\documentclass[%
 reprint,
superscriptaddress,
 amsmath,amssymb,
 aps,
prx]{revtex4-2}
\usepackage{graphicx}
\usepackage{dcolumn}
\usepackage{bm}
\usepackage{physics}
\usepackage{hyperref}
\usepackage{algorithm}
\usepackage{algpseudocode}
\usepackage[breakable,skins]{tcolorbox}
\usepackage[]{amsthm}
\usepackage{enumitem}

\renewcommand{\eqref}[1]{Eq.~(\ref{#1})}

\newcommand{\Frac}{\mathrm{Frac}}

\newtheorem{thm}{Theorem}
\newtheorem{col}{Corollary}
\newtheorem{prop}{Proposition}

\newcommand{\ketx}[1]{|{#1}\rangle}
\newcommand{\brkt}[2]{\langle{#1}|{#2}\rangle}
\newcommand{\bopk}[3]{\langle{#1}|{#2}|{#3}\rangle}

\newcommand{\figref}[1]{Fig.~\ref{#1}}

\usepackage{color}
\definecolor{blue}{rgb}{0,0.2,1}

\definecolor{red}{rgb}{0.9,0,0}

\newcommand{\past}[1]{\overleftarrow{#1}}
\newcommand{\fut}[1]{\overrightarrow{#1}}
\newcommand{\pastfut}[1]{\overleftrightarrow{#1}}

\begin{document}

\title{Dimension reduction in quantum sampling of stochastic processes}

\author{Chengran Yang}
\email{yangchengran92@gmail.com}
\affiliation{Centre for Quantum Technologies, National University of Singapore, 3 Science Drive 2, Singapore 117543}

\author{Marta Florido-Llinàs}
\email{marta.florido.llinas@mpq.mpg.de}
\affiliation{Max-Planck-Institut für Quantenoptik, Hans-Kopfermann-Straße 1, D-85748 Garching, Germany}
\affiliation{Department of Mathematics, Imperial College London, London SW7 2AZ, United Kingdom}

\author{Mile Gu}
\email{mgu@quantumcomplexity.org}
\affiliation{Nanyang Quantum Hub, School of Physical and Mathematical Sciences, Nanyang Technological University, Singapore 637371}
\affiliation{Centre for Quantum Technologies, National University of Singapore, 3 Science Drive 2, Singapore 117543}

\author{Thomas J. Elliott}
\email{physics@tjelliott.net}
\affiliation{Department of Physics \& Astronomy, University of Manchester, Manchester M13 9PL, United Kingdom}
\affiliation{Department of Mathematics, University of Manchester, Manchester M13 9PL, United Kingdom}
\affiliation{Department of Mathematics, Imperial College London, London SW7 2AZ, United Kingdom}

\begin{abstract}
Quantum technologies offer a promising route to the efficient sampling and analysis of stochastic processes, with potential applications across the sciences. Such quantum advantages rely on the preparation of a quantum sample state of the stochastic process, which requires a memory system to propagate correlations between the past and future of the process. Here, we introduce a method of lossy quantum dimension reduction that allows this memory to be compressed, not just beyond classical limits, but also beyond current state-of-the-art quantum stochastic sampling approaches. We investigate the trade-off between the saving in memory resources from this compression, and the distortion it introduces. We show that our approach can be highly effective in low distortion compression of both Markovian and strongly non-Markovian processes alike. We further discuss the application of our results to quantum stochastic modelling more broadly.
\end{abstract}

\maketitle


\section{Introduction}

Complex stochastic processes abound across the sciences, from evolutionary biology and chemistry~\cite{kulp1996generalized, grimm2005pattern, gillespie2007stochastic, wilkinson2009stochastic, smouse2010stochastic}, through geophysics and astrophysics~\cite{garavaglia2011earthquake, bartlett2022assessing}, to financial markets~\cite{arthur1999complexity, bulla2006stylized, park2007complexity, yang2008increasing}, traffic modelling~\cite{maerivoet2005cellular}, and natural language processing~\cite{levinson1986continuously, Rabiner89}. Given their pivotal role in these fields, it is vital that we can effectively simulate, analyse, and understand stochastic processes. However, the number of possible trajectories that such processes can explore generically grows exponentially over time, limiting the horizon over which we can feasibly study their behaviour. This makes tools and techniques that mitigate this growth in complexity of critical value.

Monte Carlo methods~\cite{hastings1970monte, gilks1995markov, rubinstein2016simulation} present such a technique. They use generative models to reduce computational resources by sampling from the process one trajectory at a time, and then average over many such sampled trajectories to estimate the expected values of properties of the process. The considerable successes of such techniques notwithstanding, they still suffer from certain drawbacks, such as the need for a (typically large) memory system to carry the information propagated in correlations over time in the process.

Quantum information processing provides a further route to efficient sampling, modelling, and analysis of stochastic processes. Known quantum advantages include quadratic speed-ups in analysing properties such as characteristic functions~\cite{blank2021quantum}, pricing options~\cite{rebentrost2018quantum, woerner2019quantum,stamatopoulos2020option}, enhanced expressivity~\cite{gao2022enhancing}, and significant reductions in the memory required by models~\cite{elliott2020extreme, elliott2021quantum, elliott2022quantum}. These advantages involve the preparation of quantum sample states (`\emph{q-samples}') that comprise of all possible strings of outputs in (weighted) superposition~\cite{schuld2018quantum}. Unlike other quantum approaches to superposing classical data states -- such as qRAM~\cite{giovannetti2008quantuma} and variational state preparation~\cite{chowdhury2020variational} -- the computational complexity of assembling q-samples of stochastic process trajectories need not grow exponentially; they can be constructed through a local, recurrent circuit structure prescribed by the aforementioned quantum stochastic models~\cite{binder2018practical, liu2019optimal}.

\begin{figure}
    \includegraphics[width=0.8\linewidth]{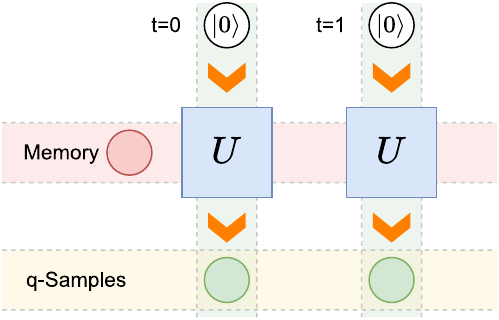}
    \caption{Recurrent quantum circuit for locally generating quantum sample states (q-samples). For each timestep a memory system is coupled with a blank ancilla; following an interaction $U$ the ancilla is entangled with the previous timesteps of the q-sample to yield the appropriate marginal q-sample state, whilst the memory is passed forward to interact with the blank ancilla for the subsequent timestep. This sequentially generates the full q-sample state for any desired number of timesteps.
    }
    \label{fig:qcirc}
\end{figure}

This recurrent circuit is illustrated in \figref{fig:qcirc}. It works by propagating a memory system -- which carries information about the correlations between the past and future of the process -- to interact with blank ancillae and sequentially assemble the q-sample state timestep by timestep. The circuit depth -- and number of ancillae qubits -- thus scales only linearly with the number of timesteps. However, as complex stochastic processes often exhibit strong temporal correlations, the requisite dimension of the memory system can grow quite large. We here explore the potential to significantly compress this memory by introducing distortion in the q-sample.

Specifically, we develop a systematic approach to \emph{lossy quantum dimension reduction}, whereby given a target q-sample circuit we determine a new circuit of fixed memory dimension that assembles an approximation to the original q-sample. Our approach is inspired by matrix product state (MPS) truncation techniques~\cite{orus2014practical, vanhecke2021tangent}, and exploits a correspondence between MPSs and q-samples~\cite{yang2018matrix}. We show that this approach yields high-fidelity approximations to q-samples corresponding to both Markovian and highly non-Markovian (i.e., strong temporal correlations) stochastic processes. We further discuss the applications of this approach in stochastic modelling~\cite{crutchfield2012between}, demonstrating that we can achieve significant compression beyond both current state-of-the-art lossless quantum dimension reduction for quantum models, and lossy compression for classical models, whilst retaining highly-accurate reconstruction of the output statistics.


\section{Framework}

\subsection{Stochastic processes and models}

A \emph{stochastic process} describes a sequence of events that are governed by a joint probability distribution~\cite{khintchine1934korrelationstheorie}. For discrete-time, discrete-event stochastic processes, at each timestep $t$ the corresponding event $x_t\in\mathcal{X}$ is described by a random variable $X_t$, drawn from a joint distribution $P(\ldots,X_t,X_t+1,\ldots)$. Sequences of events are typically correlated, and we use the shorthand $x_{t:t'}$ to denote the contiguous sequence $x_t,x_{t+1},\ldots,x_{t'-1}$. Here we consider stationary (time-invariant) stochastic processes, such that $t\in\mathbb{Z}$, and all marginal distributions of any length are shift invariant (i.e., $P(X_{0:L})=P(X_{m:m+L})\forall m\in\mathbb{Z}, L\in\mathbb{N}$. We can thus without loss of generality define $t=0$ to be the present timestep, and respectively define the full past and future event sequences as $\past{x}:=x_{-\infty:0}$ and $\fut{x}:=x_{0:\infty}$.

Such stochastic processes can be generated sequentially by a \emph{model}, consisting of an encoding function $f:\past{\mathcal{X}}\to\mathcal{M}$ that maps from sequences of past observations to a set of memory states $\mathcal{M}$, and an update function $\Lambda:\mathcal{M}\to\mathcal{X}\times\mathcal{M}$ that acts on the current memory state to (stochastically) produce the next output and update the memory state accordingly. A special class of models is that of so-called causal, or unifilar, models, where $f$ is a deterministic mapping, and $\Lambda$ has a deterministic update rule for the memory given the input memory state and output event~\cite{shalizi2001computational}. Amongst all classical unifilar models that reproduce the exact statistics of the process, the provably-memory-minimal (both in terms of information stored and memory dimension) is the $\varepsilon$-\emph{machine}, prescribed by an encoding function $f_\varepsilon$ that satisfies $f_\varepsilon(\past{x})=f_\varepsilon(\past{x}')\Leftrightarrow P(\fut{X}|\past{x})=P(\fut{X}|\past{x}')$~\cite{shalizi2001computational}. The corresponding memory states $s\in\mathcal{S}$ are referred to as the \emph{causal states} of the process~\cite{crutchfield1989inferring}.

\subsection{Quantum models and q-samples}

Curiously, it is possible to push the memory cost below classical limits when using quantum encoding and update functions, even though we are considering classical stochastic processes~\cite{gu2012quantum, mahoney2016occam, binder2018practical, liu2019optimal, ghafari2019dimensional}. Such quantum models use an encoding function that maps to quantum states, and the update function is a quantum channel. The current state-of-the-art unifilar quantum models are based on an encoding function $f_q$ that, like the classically-minimal encoding function $f_\epsilon$, maps pasts to the same memory state iff they belong to the same causal state~\cite{liu2019optimal}; we denote the quantum memory state corresponding to causal state $s_j$ as $\ket{\sigma_j}$. These memory states are defined implicitly by the update function, which can be expressed in terms of a unitary operator $U$:
\begin{equation}
\label{eq:unitary}
U\ket{\sigma_j}\ket{0}=\sum_x\sqrt{P(x|j)}e^{i\varphi_{xj}}\ket{\sigma_{\lambda(x,j)}}\ket{x},
\end{equation}
where $P(x|j)$ is the probability that the next symbol is $x$ given we are in causal state $s_j$, $\lambda(x,j)$ is an update rule that outputs the updated memory state label, and $\{\varphi_{xj}\}$ are a set of real numbers. By measuring the second subsystem after this evolution, we obtain the output event for the next timestep, and leave the memory state in the first subsystem, appropriately updated. Repeating this procedure multiple times with fresh ancillae initialised in $\ket{0}$, the statistics can be replicated for any desired number of timesteps (see \figref{fig:qcirc}). This expression defines a valid evolution for any choice of $\{\varphi_{xj}\}$, though each choice defines a different set of memory states -- and thus yields different memory costs~\cite{liu2019optimal}. These parameters thus constitute a tunable element of the model; nevertheless, any choice will yield a memory cost that does not sit higher than that of the classical minimal model, neither in terms of information stored or dimension.

Suppose we do not measure the output subsystem at each timestep. If we preserve this at every timestep, then we will build a large entangled state that is a weighted superposition over all possible trajectories~\cite{binder2018practical}. If we set all $\{\varphi_{xj}\}$ to 0, then after $L$ steps we obtain the state $\sum_{x_{0:L}}\sqrt{P(x_{0:L}|j)}\ket{x_{0:L}}\ket{\sigma_{\lambda(x_{0:L},j)}}$. We can then perform an operation on the final subsystem, conditioned on knowledge of the initial state label $j$ and controlled by the other subsystems, to deterministically set it to $\ket{0}$. This leaves us with the corresponding $L$-step q-sample:
\begin{equation}
\label{eq:qsample}
\ket{P(X_{0:L}|j)}:=\sum_{x_{0:L}}\sqrt{P(x_{0:L}|j)}\ket{x_{0:L}}.
\end{equation}
As discussed above, such q-samples are a resource for achieving quantum advantages in sampling tasks, particularly speed-ups. We have been able to exploit the serial nature of the data to prescribe a recursive circuit that assembles the q-sample to any desired length $L$ with a circuit depth that grows only linearly with $L$.

However, there is still a cost that grows with the complexity of the process -- the size of the memory system. In practical terms, this can be quantified by the number of qubits required for the memory, i.e., $D_q:=\log_2(d)$, where $d$ is the dimension of the memory. This dimension is upper-bounded by the number of memory states, though can be lower when they exhibit linear dependencies~\cite{liu2019optimal, elliott2020extreme}. Also of relevance is the amount of information stored within the memory, i.e., the von Neumann entropy of the memory states. That is, $C_q:=S(\rho)$, where $\rho:=\sum_jP(j)\ket{\sigma_j}\bra{\sigma_j}$ is the steady-state of the memory system. Clearly, we must have that $C_q\leq D_q$; for many stochastic processes we even find that $C_q\ll D_q$~\cite{aghamohammadi2016extreme, Garner2017Unbounded, elliott2018superior, elliott2019memory}. This indicates that much of the memory capacity is under-utilised given the amount of information that must be stored. In turn, this suggests that it may be possible to drastically reduce the memory dimension whilst discarding very little information, and thus assemble high-fidelity approximations to q-samples with drastically reduced memory size.


\subsection{MPS representation of q-samples}

\begin{figure}[htbp]
    \centering
    \includegraphics[width =0.8\linewidth]{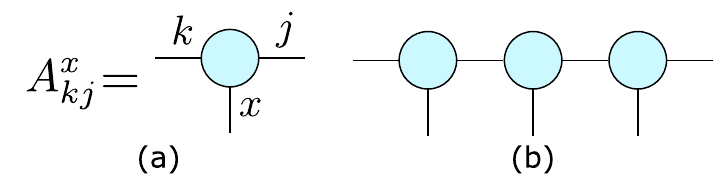}
    \caption{Diagrammatic representation of matrix product states. (a) Each site of a 1D quantum chain can be described by a site matrix $A^x_{jk}$, where $x$ is the physical index corresponding to the local state, and $j$ and $k$ are bond indices carrying correlations between sites. (b) The full MPS is constructed by joining together these site matrices by multipyling (`contracting') over the bond indices.
    }
    \label{fig:mps}
\end{figure}

MPSs are a powerful tool for finding efficient representations of one-dimensional (1D) quantum systems and simulating their evolution\cite{fannes1992finitely, Schon2005, perez2006matrix, Orus2007, orus2014practical}. A general 1D quantum chain can be brought into MPS form by a series of Schmidt decompositions bipartitioning each neighbouring pair of sites of the chain, resulting in a series of site matrices $\{A^{x,[l]}_{kj}\}$, where $l$ represents the site index, $x$ the label of the corresponding local basis state on that site, and $j$ and $k$ are the so-called `bond' indices between sites. For a 1D quantum chain with state $\sum_{x_1,\ldots,x_L}a_{x_1,\ldots,x_L}\ket{x_1,\ldots,x_L}$, we then have that $\Tr(A^{x_1,[1]}\ldots A^{x_L,[L]})=a_{x_1,\ldots,x_L}$. For infinite, translationally-invariant chains the site matrices are homogenous, and the site index can be dropped (see \figref{fig:mps} for a diagrammatic representation); these are referred to as infinite MPS (iMPS)~\cite{orus2014practical}. To determine amplitudes for contiguous subregions of the chain, the trace can be replaced by appropriate boundary vectors. See Appendix~\ref{app:mps} for further details on MPS, including discussion of their canonical forms.

Given that q-samples (see \eqref{eq:qsample}) are a form of 1D quantum state, they can be recast in an MPS representation~\cite{Schon2005}. For q-samples corresponding to the output of a stationary stochastic process, a valid choice of site matrices is given by
\begin{equation}
\label{eq:mpsrep}
A^x_{kj}=\sqrt{P(x|j)}\delta_{\lambda(x,j),k},
\end{equation}
where $\delta_{jk}$ is the Kronecker delta~\cite{yang2018matrix}. For q-samples generated by models with non-zero $\phi_{xj}$, these can be constructed by appending the appropriate phase factors to the elements in \eqref{eq:mpsrep}~\cite{martathesis}. Moreover, the left canonical form of this MPS has site matrices corresponding to the Kraus operators of the channel acting on the memory system in \eqref{eq:unitary}, and the associated Schmidt coefficients for the bond correspond to the spectrum of the steady-state of the memory system used to assemble the q-sample -- and thus readily reveal the associated memory costs~\cite{yang2018matrix}.

\section{Quantum dimension reduction for q-samples}

A key utility of the MPS representation of 1D quantum chains is that they present a means of identifying efficient representations of the underlying quantum state~\cite{Verstraete2006}. While in principle the dimension of the bond indices can grow exponentially with the length of the chain, for chains that correspond to the ground state of gapped local Hamiltonians the number of non-neglible Schmidt coefficients in each bond tends towards a constant. That is, the number of bond indices carrying non-negligible correlations between sites remains finite even in the thermodynamic limit. This is because such Hamiltonians satisfy an `area law', where the entanglement in the ground state is non-extensive. Crucially, we can then discard the remaining Schmidt coefficients and construct a close approximation of the MPS with finite bond dimension, leading to an efficient, high-fidelity approximate representation of the quantum state.

We apply this approach to q-samples and their assembly by recurrent quantum circuits. That $C_q$, the information stored by the memory generating a q-sample, is often bounded for many classes of complex stochastic processes suggests a form of `temporal area law' for stochastic process q-samples. We capitalise upon this to identify high-fidelity approximations to q-samples with memory size $\tilde{D}_q\ll D_q$, and prescribe the corresponding quantum circuits to efficiently assemble them.

To quantify the accuracy of our approximate q-samples, we will use the quantum fidelity divergence rate (QFDR)~\cite{vanhecke2021tangent}. This extends the notion of quantum fidelity (i.e., state overlap) from finite-dimensional states to infinite length chains. This is because any distortion introduced in the state at each timestep will -- no matter how small -- result in the fidelity between the states of the infinite chain going to zero. The QFDR $R_F$ therefore captures the rate at which this distortion reduces the fidelity on a per timestep basis. Specifically, 
\begin{equation}
\label{eq:qfdr}
R_F\left(\ket{P},\ket{P'}\right):=-\lim_{L\to\infty} \frac{1}{2L} \log_2 F\left(\ket{P_L},\ket{P_L'}\right),
\end{equation}
where $F(\ket{\psi},\ket{\phi})=|\brkt{\psi}{\phi}|$ is the quantum fidelity between states, $\ket{P_L}$ is shorthand for the $L$-length q-sample $\ket{P(X_{0:L})}$, and $\ket{P}$ the $L\to\infty$ limit of this. Given iMPS representations $\{A^x\}$ and $\{{A'}^x\}$ of $\ket{P}$ and $\ket{P'}$ respectively, this quantity is efficiently calculable, and can be expressed in terms of the leading eigenvector of the matrix $\sum_x A^x\otimes({A'}^x)^*$, where $A^*$ is the element-wise complex conjugate of $A$~\cite{yang2020measures, vanhecke2021tangent}. Thus, as a measure of distortion we consider the QFDR between a q-sample and its approximation.


\subsection{Theoretical bounds}

The direct form of truncation for MPSs is to reduce them to their canonical form, keep the $\tilde{d}$ largest Schmidt coefficients at each bond (where $\tilde{d}$ is the desired final bond size), discard the rest, and then uniformly rescale the remaining Schmidt coefficients such that the resulting MPS is appropriately normalised~\cite{vidal2003efficient, daley2004time, white2004real, Orus2007}. For the case of q-samples, described by iMPS, we can upper-bound the QFDR $R_F$ in terms of the magnitude of the discarded Schmidt coefficients.

\begin{thm}\label{thm:compbound}
Consider a q-sample $\ket{P}$ for which the iMPS representation has $d$ Schmidt coefficients $\{\lambda_j\}$ labelled in decreasing order. For any truncated dimension $\tilde{d}$, there always exists a q-sample $\ketx{\tilde{P}}$ that can be constructed sequentially with a memory of at most dimension $\tilde{d}$ that satisfies
\begin{equation}
R_F(\ket{P},\ketx{\tilde{P}})\leq\frac{\epsilon_{\tilde{d}}}{2(1-L\epsilon_{\tilde{d}})\ln 2}+O\left(\frac{1}{L}\right)
\end{equation}
for any $L\in \mathbb{N}$, where $\epsilon_{\tilde{d}}:=\sum_{k=\tilde{d}+1}^d\lambda_k$.
\end{thm}
The proof is given in Appendix~\ref{app:compbound}. Moreover, by setting $L\sim 1/2\epsilon_{\tilde{d}}$ we obtain that $R_F(\ket{P},\ketx{\tilde{P}})\leq O(\epsilon_{\tilde{d}})$. Thus, we see that the QFDR grows only linearly with the sum of discarded Schmidt coefficients. By choosing $\tilde{d}$ such that $\epsilon_{\tilde{d}}$ is sufficiently small, we can therefore control the distortion between the target q-sample and its reduced memory approximation. 

However, this does not by itself carry any guarantees that such a $\tilde{d}$ is much smaller than the original memory dimension $d$. In the following theorem and corollary we establish a relationship between the information cost of the q-sample, the truncated memory dimension, and the size of the sum of discarded Schmidt coefficients.

\begin{thm}\label{thm:truncboundmps}
Given an iMPS with $d\geq 3$ non-zero Schmidt coefficients, and its approximation by truncation to the largest $\tilde{d}\geq 3$ Schmidt coefficients, the sum of the truncated Schmidt coefficients $\epsilon_{\tilde{d}}$ is bounded as follows
\begin{equation}
\label{eq:truncboundmps}
\epsilon_{\tilde{d}}\leq \frac{H(\lambda)}{\frac{\tilde{d}-2}{d-\tilde{d}}\log_2(d-\tilde{d})+\log_2(\tilde{d})},
\end{equation}
where $H(\lambda)$ is the Shannon entropy of the Schmidt coefficients.
\end{thm}
The proof is given in Appendix~\ref{app:truncboundmps}. By recasting this theorem in terms of q-samples and by further loosening \eqref{eq:truncboundmps} we obtain the following corollary:
\begin{col}
\label{col:truncbound}
Consider a q-sample $\ket{P}$ for which the iMPS representation has $d\geq 3$ non-zero Schmidt coefficients and can be constructed sequentially by a memory with an information cost $C_q$. Consider also an approximation to this q-sample $\ketx{\tilde{P}}$ with truncated dimension $\tilde{d}\geq 3$ formed by truncating the $d-\tilde{d}$ smallest Schmidt coefficients of $\ket{P}$. The sum of the truncated Schmidt coefficients $\epsilon_{\tilde{d}}$ satisfies
\begin{equation}
\epsilon_{\tilde{d}}\leq \frac{C_q}{\tilde{D}_q},
\end{equation}
where $\tilde{D}_q:=\log_2(\tilde{d})$.
\end{col}
Further, by combining this with Theorem~\ref{thm:compbound}, we have that for any q-sample $\ket{P}$ of memory dimension $d\geq 3$ and information cost $C_q$, there exists a q-sample $\ketx{\tilde{P}}$ of memory dimension $\tilde{d}$ that approximates $\ket{P}$ satisfying
\begin{equation}
R_F(\ket{P},\ketx{\tilde{P}})\leq O\left(\frac{C_q}{\tilde{D}_q}\right),
\end{equation}
where $\tilde{D}_q=\log_2(\tilde{d})$. This indicates that if $C_q\ll D_q$, there is significant scope for truncating the memory dimension without introducing significant error, affirming our earlier intuition.


\subsection{Computational Approach}

To carry out our quantum dimension reduction for q-sample construction, we use the \emph{variational truncation approach} for iMPS, based on tangent-space methods~\cite{zauner2018variational, vanhecke2021tangent, vanderstraeten2019tangent}. Given an injective iMPS~\footnote{An injective iMPS is one for which the transfer matrix $\sum_x A^x\otimes (A^x)^*$ has a non-degenerate leading eigenvalue of 1. The q-sample of an ergodic stochastic process corresponds to an injective iMPS representation.} $\{A^x\}$ of bond dimension $d$, this approach seeks to find an injective iMPS $\{\tilde{A}^x\}$ of dimension $\tilde{d}$ of maximum fidelity with the target iMPS. This is equivalent to finding the iMPS representation of a q-sample $\ketx{\tilde{P}}$ with memory cost $\tilde{D}_q=\log_2(\tilde{d})$ that minimises the QFDR with respect to a target q-sample $\ket{P}$ with memory cost $D_q=\log_2(d)$. We then use the iMPS representation of the approximating q-sample to prescribe a circuit that can be used to construct it. The procedure is summarised in the box ``\emph{Variationally-truncated q-samples}".

This computational approach involves several MPS gauges: the left and right canonical forms $A_l$ and $A_r$, and the mixed gauge $\{A_c, C\}$, explained in detail in Appendix~\ref{app:mps}. Specifically, it relies on the fact that the minimality condition for the QFDR can be expressed in terms of a tangent-space projector of the iMPS, which has an explicit representation in the mixed-gauge canonical form of the iMPS. Here we outline the computational steps; further details on how and why this approach to iMPS truncation works can be found in~\cite{vanhecke2021tangent}.

For the initial stage, the first step involves randomly generating an initial ansatz injective iMPS for $\tilde{A}$ of bond dimension $\tilde{d}$, and determining its associated left and right canonical forms and mixed gauge tensors. Similarly, we use \eqref{eq:mpsrep} to obtain the iMPS representation of the target q-sample, and calculate the associated left and right canonical form. 

In the second step we then construct the left and right mixed transfer matrices, defined as follows:
\begin{equation}
\label{eq:mixedtrans}
    \bar{\mathbb{E}}_l =  \sum_x A_l^x\otimes (\tilde{A}_l^x)^{*} \quad \bar{\mathbb{E}}_r =  \sum_x A_r^x\otimes (\tilde{A}_r^x)^{*}.
\end{equation}
Subsequently, we perform an eigendecomposition on these matrices to find their leading eigenvalue $\eta$, and associated left (for $\bar{\mathbb{E}}_l$) and right (for $\bar{\mathbb{E}}_l$) leading eigenvectors $G_l$ and $G_r$ respectively.

\begin{figure}
    \centering
    \includegraphics[width=0.8\linewidth]{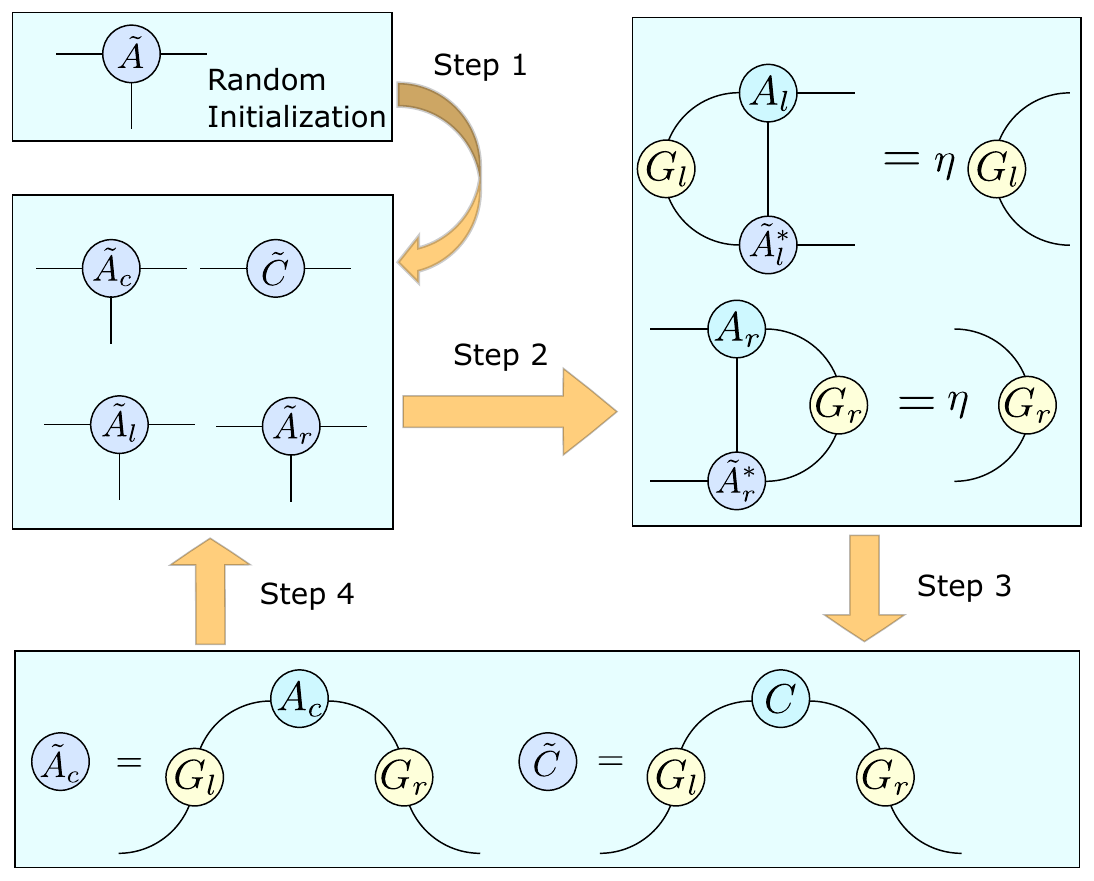}
    \caption{Variational truncation approach for iMPS. From an initial ansatz iMPS an iterative approach is employed where at each iteration a new ansatz is generated that is closer (i.e., has lower QFDR) to the target iMPS, and used as the input for the next iteration. The steps are repeated until the error falls below a predefined threshold.
    }
    \label{fig:vtrunc}
\end{figure}

In the third step we use these eigenvectors to update our truncated ansatz iMPS, according to
\begin{equation}
    \tilde{A}_c \to G_lA_cG_r,\quad \tilde{C} \to G_lCG_r.
\end{equation}
Then, as the fourth step we calculate the left and right canonical form of this new ansatz. We feed this back in to step 2 and iterate until the error $\Delta$ falls below a predetermined threshold, where
\begin{equation}
\label{eq:error}
    \Delta := \norm*{\tilde{A}_c/\eta - \tilde{A}_l\tilde{C}}.
\end{equation}
These steps are illustrated diagrammatically in \figref{fig:vtrunc}.

We remark that in precursor work we compared this approach also with the direct truncation approach, i.e., where only the $\tilde{d}$ largest Schmidt values are kept and the rest discarded, and found that the variational approach achieved stronger performance, and was more stable due to explicitly preserving injectivity of the MPS~\cite{martathesis}. One could also consider a two-step approach, where the direct truncation is fed into the variational approach as the initial ansatz.

The final stage is to turn this reduced-dimension iMPS $\tilde{A}$ into a quantum circuit that assembles the approximating q-sample. To do this, we find the left canonical form $\tilde{A}_l$ of $\tilde{A}$, which satisfies $\sum_x ({\tilde{A}_l^x})^\dagger \tilde{A}^x_l=\mathbb{I}$. That is, the $\tilde{A}^x_l$ define a valid set of Kraus operators for a completely-positive, trace-preserving (CPTP) quantum channel $\mathcal{E}(\cdot)=\sum_x\tilde{A}^x_l \cdot ({\tilde{A}^x_l})^\dagger$ acting on a system of dimension $\tilde{d}$~\cite{Schon2005, yang2018matrix}. Taking this system to be the memory used to assemble the q-sample, we can dilate this into a unitary operator $\tilde{U}$ such that for all basis states $\{\ket{j}\}$ of the memory system,
\begin{equation}
\label{eq:truncu}
\tilde{U}\ket{j}\ket{0}=\sum_x (\tilde{A}^x_l\ket{j})\ket{x}.
\end{equation}
The remaining elements of $\tilde{U}$ which act on initial states of the ancilla orthogonal to $\ket{0}$ can be assigned arbitrarily, provided they preserve the unitarity of $\tilde{U}$~\footnote{This can be achieved using a Gram-Schmidt procedure~\cite{binder2018practical}.}. It can be shown that recursively applying $\tilde{U}$ with a fresh ancilla at each timestep will then assemble the q-sample $\ketx{\tilde{P}}$ associated with the iMPS $\tilde{A}$, with the initial condition dependent on the initial memory state. 

We can use the left canonical form to determine the appropriate state to initialise the memory in given some specified conditional past. For a past $x_{-L:0}$, we prepare the memory in $\tilde{f}_q(x_{-L:0}):=\rho_{x_{-L:0}}\propto \tilde{A}_l^{x_{-L:0}}\rho ({\tilde{A}^{x_{-L:0}}_l})^\dagger$, where $\rho$ is the fixed point of $\mathcal{E}$, $A^{x_{0:L}}:=A^{x_L-1}\ldots A^{x_1}A^{x_0}$, and the proportionality is such that the state is appropriately normalised. To assemble the q-sample with no conditioning on the past, the memory should be initialised in $\rho$, i.e., the steady-state, or purification thereof.

\begin{figure}
    \vspace*{0.5em}
    \begin{tcolorbox}[skin=enhanced]
     {\bf Variationally-truncated q-samples}
    
     \vspace*{0.5em}
     {\bf Inputs:}\\
     \vspace*{0.25em}
     \begin{tabular}{rcl}
        $\mathcal{P}$ & -- & the target process.\\
        $\tilde{d}$  &  --  & the desired truncated memory dimension. \\
        $\Delta_\mathrm{thresh}$ & -- & the desired error tolerance.
        \end{tabular}
    
     \vspace*{0.5em}
     {\bf Outputs:}\\
     \vspace*{0.25em}
     \begin{tabular}{rcl}
        $\tilde{U}$  &  --  & the unitary dynamics to generate the truncated\\ & & approximate q-sample $\ketx{\tilde{P}}$. \\
        $\tilde{f}_q$ & -- & the encoding map from pasts to truncated\\ & & memory states.
     \end{tabular}	
    
     \vspace*{0.5em}
     {\bf Algorithm:}
        \vspace*{-0.5em}
        \begin{enumerate}
             \item Obtain an iMPS representation of $\ket{P}$, the \mbox{q-sample} of $\mathcal{P}$, according to~\eqref{eq:mpsrep} and calculate the associated left and right canonical forms $A_l$ and $A_r$.
        
            \item Determine the iMPS representation $\tilde{A}$ of the truncated q-sample $\ketx{\tilde{P}}$:
            \vspace*{-0.5em}
            \begin{enumerate}
            \item Randomly initialize tensor $\tilde{A}$ with dimension $\tilde{d}$.
            \item Compute the left and right canonical forms of $\tilde{A}$, $\tilde{A}_l$ and $\tilde{A}_r$, and the mixed gauge $\{\tilde{A}_c,\tilde{C}\}$.
            \item Compute leading eigenvalue and eigenvectors $\eta, G_r, G_r$ according to $G_l\bar{\mathbb{E}}_l=\eta G_l$ and $\bar{\mathbb{E}}_rG_r=\eta G_r$, using the definitions~\eqref{eq:mixedtrans}.
            \item Update $\tilde{A}_c\to G_lA_cG_r, \tilde{C} \to G_lCG_r$.
            \item Evaluate $\Delta$ using \eqref{eq:error}.
            \item \textbf{Repeat} (b)-(e) \textbf{Until} {$\Delta<\Delta_\mathrm{thresh}$.}
            \end{enumerate}
            
            \item Determine $\tilde{U}$:
            \begin{enumerate}
            \item Calculate the left canonical form $\tilde{A}_l^x$ of $\tilde{A}$.
            \item Define the associated columns of $\tilde{U}$ according to \eqref{eq:truncu}.
            \item Use the Gram-Schmidt procedure to fill the remaining columns of $\tilde{U}$, preserving unitarity.
            \end{enumerate}

            \item Determine $\tilde{f}_q$:
            \vspace*{-0.5em}
            \begin{enumerate}
            \item Determine the fixed point $\rho$ of \mbox{$\mathcal{E}(\cdot)=\sum_x\tilde{A}^x_l \cdot ({\tilde{A}^x_l})^\dagger$}.
	    \item Define $\tilde{f}_q$ according to \mbox{$\tilde{f}_q(x_{-L:0}):=\dfrac{\tilde{A}_l^{x_{-L:0}}\rho ({\tilde{A}^{x_{-L:0}}_l})^\dagger}{\Tr\left(\tilde{A}_l^{x_{-L:0}}\rho ({\tilde{A}^{x_{-L:0}}_l})^\dagger\right)}$}.
            \end{enumerate}
        \end{enumerate}
    \end{tcolorbox}
\end{figure}


\section{Examples}
\label{sec:examples}

We demonstrate the application of our quantum dimension reduction with two illustrative examples that capture two different extremes of stochastic processes with high memory cost. The first are cyclic random walks, which are Markovian processes with large event alphabets~\cite{Garner2017Unbounded}. The second is the sequential read-out of a Dyson-Ising spin chain, which manifests a highly non-Markovian stochastic process~\cite{aghamohammadi2016extreme}. In both cases, our algorithm is able to identify new quantum circuits that reconstruct the ideal q-sample with low QFDR, and yet have drastically reduced memory cost compared to the original circuit.


\subsection{Cyclic random walks}

\begin{figure}
    \centering
    \includegraphics[width=0.6\linewidth]{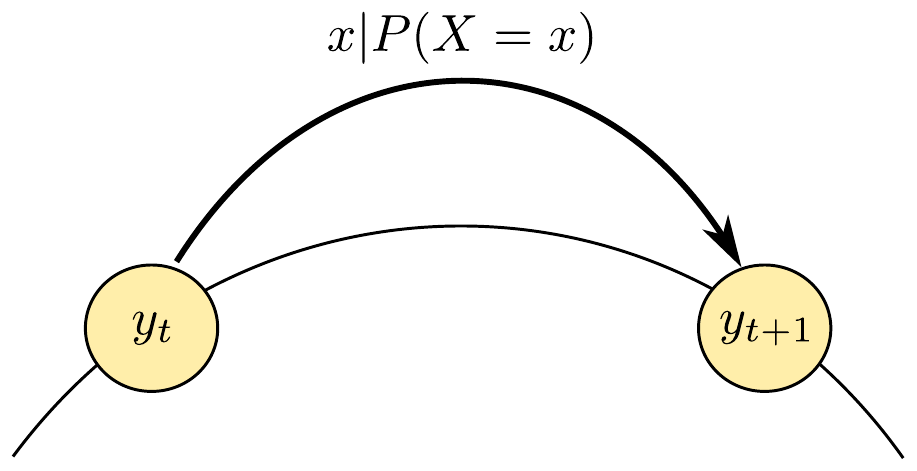}
    \caption{Cyclic random walks. At each timestep, the walker shifts position around the ring by some stochastic variable $X$; $y_{t+1}=\Frac(y_t+x)$.
    }
    \label{fig:crwalks}
\end{figure}

A cyclic random walk describes the stochastic motion of a particle on a ring, as depicted in \figref{fig:crwalks}~\cite{Garner2017Unbounded}. Without loss-of-generality, the length of the ring can be taken to be one unit, such that the position of the walker is given by $y\in[0,1)$, where due to the periodicity of the ring positions outside of this range can be mapped back within it by taking the fractional component (e.g., $\Frac(5.46)=0.46$). At each timestep the walker shifts its position by a stochastic increment $x$, i.e.,
\begin{equation}
y_{t+1}=\Frac(y_t+x),
\end{equation}
where $x$ is drawn from some distribution $Q(X)$.

It is clear that this constitutes a Markovian process, as the next position of the walker depends only on its current position and a memoryless stochastic variable. However, exact specification of the current position of the walker is given by a real number, and so to avoid this divergence in memory cost we consider approximations based on $n$-bit discretisations of the current position. The natural approach to this discretisation is to divide the ring into $N=2^n$ segments of equal length, labelled $I_k$, i.e., $I_k=\left\{y:\left |y-k/N \right | <1/2N\right\}$. We then describe the approximate position of the walker according to the interval its position belongs to, and denote this state as $S_k$. The corresponding Markov chain is then found by integrating $Q(X)$ over these regions to obtain a distribution $P(S_k|S_j)=P(Y_{t+1}\in I_k|y_t\in I_j)$. It can be shown that the $\{S_k\}$ correspond to the causal states of the process~\cite{Garner2017Unbounded}.

Since each interval carries equal probability mass, this means that for general $Q(X)$ the memory minimal unifilar classical model not only requires $N$ memory states, but also has an information cost of $\log_2(N)$. In contrast, it has been shown that for several families of distribution for $Q(X)$, the quantum information cost $C_q$ can remain finite as $N\to\infty$, even while $d_q$ diverges as $N$, akin to the classical model~\cite{Garner2017Unbounded}. Such scenarios present an excellent opportunity to test our lossy dimension reduction for constructing q-samples of these processes, and demonstrate Theorem~\ref{thm:compbound} in action.

\begin{figure}
    \centering
    \includegraphics[width=\linewidth]{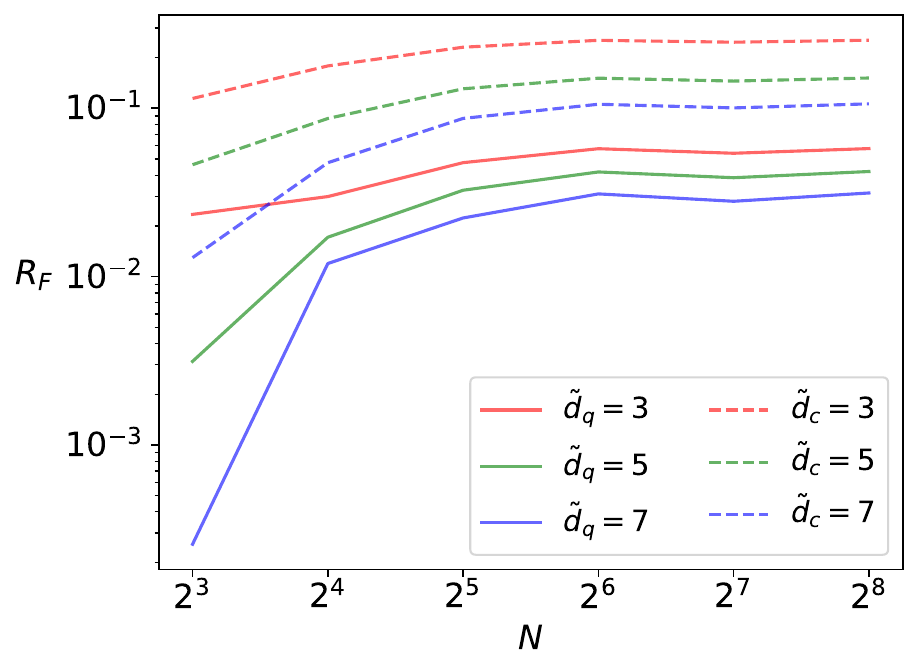}
    \includegraphics[width=\linewidth]{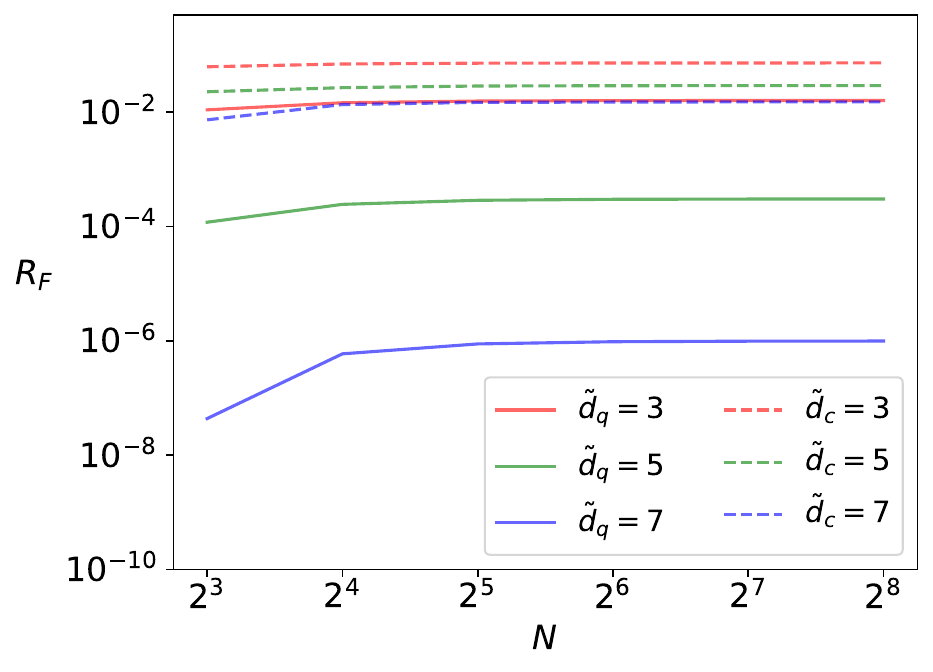}  
    \caption{
    Fidelity divergence rates for cyclic random walks with (a) uniformly- and (b) normally-distributed shifts. We plot the QFDR $R_F$ of reduced memory quantum circuits to generate q-samples of such processes. This is an upper-bound on the corresponding statistical FDR of a quantum model of the process. We also plot the statistical FDR for dimension-reduced classical models, showing inferior performance to the quantum models.
    }
    \label{fig:tcrwalksu}
\end{figure}

As a concrete example, we consider the case where $Q(X)$ is uniformly distributed over the interval $[-0.1,0.1]$, and 0 elsewhere. We examine the perfomance of our reduced memory q-sample circuits for truncated dimensions $\tilde{d}_q=\{3,5,7\}$. The corresponding QFDRs $R_F$ are plotted in \figref{fig:tcrwalksu}(a). We remark that for $N\to\infty$, $C_q\approx 3.4$ for this process, and so even with $\tilde{D}_q\leq C_q$, we can maintain a reasonably high degree of accuracy (distortion $O(10^{-2})$bits/timestep). As a further example, we carry out a similar computation for $Q(X)$ taking the form of a normal distribution with zero mean and width 0.1, finding qualitatively similar results [\figref{fig:tcrwalksu}(b)]. In this case, we find that even at $N=256$, our method can achieve distortions below $10^{-6}$bits/timestep with a memory dimension of only $\tilde{d}_q=7$ -- substantially reducing the memory cost involved in generating the associated q-sample.

\subsection{Dyson-Ising spin chains}

The Dyson-Ising spin chain generalises the Ising model to non-neighbouring interactions~\cite{baxter2016exactly,dyson1969existence}. Its dynamics are described by the (classical) Hamiltonian
\begin{equation}
H_{\mathrm{DI}}=\sum_{j,k}J(j,k) \sigma_j\sigma_k,
\end{equation}
where $\sigma_j$ is the spin at site $j$, taking on values $\{\pm 1\}$. The interaction takes the form $J(j,k)=J_0/|j-k|^\delta$ for some interaction strength scaling $J_0$ and decay rate $\delta$. This infinite-range interaction is then typically approximated by a finite-length truncation, where $J(j,k)$ is set to zero for $|j-k|>L$ for truncation length $L$; we denote the corresponding Hamiltonian as $H_{\mathrm{DI}}^{(L)}$.

Consider the thermal state of this Hamiltonian at temperature $T$. The probability of a given spin configuration $\pastfut{\sigma}$ is given by
\begin{equation}
P(\pastfut{\sigma})=\frac{1}{Z}e^{-\frac{H_{\mathrm{DI}}^{(L)}}{T}},
\end{equation}
where the normalisation $Z$ is the partition function. This can be viewed as a stochastic process, where we sweep sequentially along the chain spin by spin, and consider the distribution for the state of the next spin given the previous spins seen~\cite{aghamohammadi2016extreme}. It can be shown that the corresponding conditional distribution for a given site is a function of only the last $L$ spins -- it is a Markov order $L$ process. Thus, as $L$ is increased the non-Markovianity of the process increases.

Moreover, each of the $2^L$ possible configurations of these last $L$ spins gives rise to a different conditional distribution. Each such configuration thus corresponds to a causal state of the process, and so the minimal unifilar classical model requires $2^L$ memory states. Previous study of these models has shown that the classical information cost is a monotonically-increasing function of $T$, and asymptotically tends towards $L$ as $T\to\infty$~\cite{aghamohammadi2016extreme}.

Meanwhile, it can be shown that there exists a quantum model with an information cost $C_q$ of at most 1 qubit for any $L$ and $T$~\cite{aghamohammadi2016extreme}. Nevertheless, the memory states still inhabit a $2^L$-dimensional space, akin to the classical models. We therefore ask whether our approach to quantum dimension reduction can be used to find means of constructing high-fidelity q-samples of the Dyson-Ising chain process with at most 1 qubit of memory (i.e., $\tilde{d}_q=2$).

\begin{figure}
    \centering
\includegraphics[width=\linewidth]{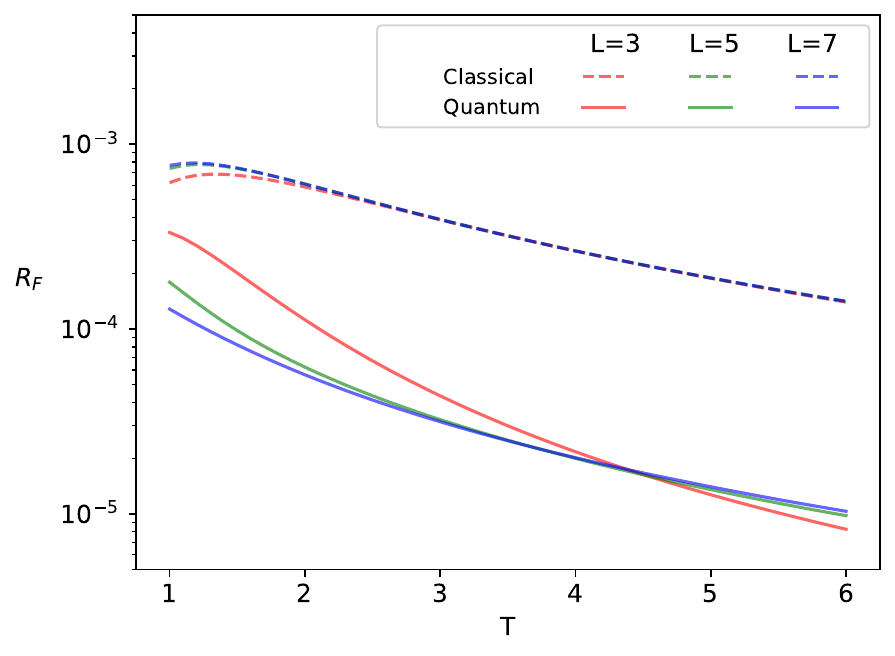}\\
\includegraphics[width=\linewidth]{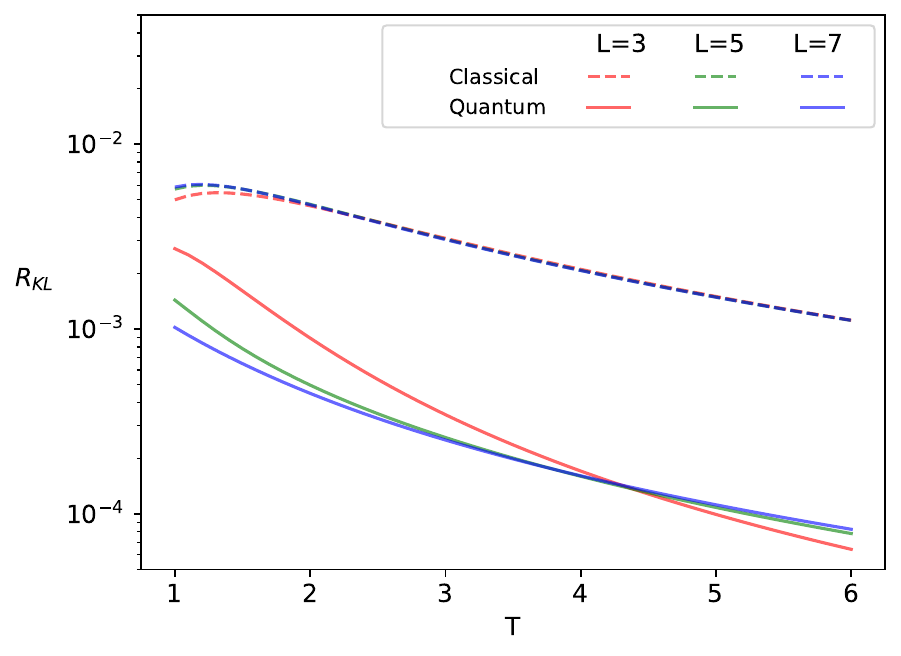}
    \caption{
    Divergence rates for 1 (qu)bit q-samples and models of the Dyson-Ising chain process. (a) QFDR $R_F$ of reduced memory quantum circuits to generate q-samples of such processes, and statistical FDR for dimension reduced classical models, showing inferior performance to the quantum models. (b) KL divergence rate $R_{\mathrm{KL}}$ of the statistics generated by the reduced dimension q-sample, and lower bound on $R_{\mathrm{KL}}$ for reduced dimension classical models. Plots shown for $J_0=1$, $\delta=2$, $L=\{3,5,7\}$.
    }
    \label{fig:tdising}
\end{figure}

Our findings answer this in the affirmative. As can be seen in \figref{fig:tdising}(a), for high $T$ our reduced memory q-samples achieve QFDRs below $10^{-3}$qubits/timestep, even as $L$ increases. Intriguingly, for low $T$, the QFDR actually \emph{decreases} with $L$, despite the increasing non-Markovianity. While we lack a rigorous argument, we suggest that this could be because the truncated q-sample circuits tend to display infinite Markov order behaviour in their statistics, and so in certain circumstances the reduced memory q-samples may more naturally fit to processes with infinite Markov order.

\section{Application to quantum stochastic modelling}

As is evident from our presentation of the framework, there is a clear connection between q-samples and quantum stochastic models. Our approach to quantum dimension reduction for sequential q-sample generation thus also corresponds to an approach for quantum dimension reduction for the memory of quantum stochastic models. Indeed, the truncated unitary dynamics $\tilde{U}$ and encoding function for truncated memory states $\tilde{f}_q$ output by our algorithm already define a reduced memory dimension quantum model; all one needs to do is measure the output system at each timestep rather than generate the whole q-sample.

How well do these truncated quantum models perform relative to classical models with truncated memory? We begin by stating a simple -- yet important -- proposition relating to compressed classical and quantum models of stochastic processes.
\begin{prop}
Given a (stationary, discrete-time, discrete-event) stochastic process, suppose there exists a classical model of dimension $\tilde{d}$ that replicates the statistics of the process to within distortion $R$, for some suitably-chosen measure of distortion. Then there exists a quantum model of dimension $\tilde{d}$ of the process with a distortion of at most $R$.
\end{prop}
This follows trivially from the fact that classical models are a subset of quantum models. An explicit means of converting any classical model into a corresponding quantum one has previously been exhibited~\cite{elliott2021memory}. Thus, given any classical model achieving a particular distortion, there exists a quantum model with the same distortion. The question remains of whether there then also exists a quantum model with \emph{lower} distortion.

For our measure of distortion, we generically encounter the same issue as we did for state fidelity -- that similarities of distributions of sequences decay to zero as the length of sequence increases. This similarly motivates the use of statistical divergence rates, which quantify the asymptotic rate at which these similarities decay~\cite{yang2020measures}. One such measure is the statistical fidelity divergence rate (statistical FDR):
\begin{equation}
R_F^{\mathrm{(S)}}(P,P'):=-\lim_{L\to\infty}\frac{1}{L}F^{(S)}(P_L,P_L'),
\end{equation}
where $F^{(S)}(p,q)=\sum_{x}\sqrt{p(x)q(x)}$ is the statistical fidelity between distributions, and $P_L$ represents the probability distribution for length $L$ sequence drawn from the stochastic process. Profitably for our purposes, the QFDR between the q-samples of these two stochastic processes also corresponds to the statistical FDR. When the q-samples have non-zero phase factors (as will typically occur in our dimension reduced approximate q-samples), the QFDR is a strict upper-bound on the statistical FDR. Thus, a low QFDR also corresponds to a low statistical FDR, and hence a good model.

Thence, we can repurpose our bounds on the performance of truncated q-samples and directly apply them to bound the performance of dimension reduced quantum stochastic models.
\begin{col}
Consider a stochastic process $P$ that can be modelled exactly by a quantum model with information cost $C_q$ and memory dimension $d$. Let the steady-state of this model have spectrum $\{\lambda_j\}$, labelled in decreasing order. Then, there exists a quantum model with memory dimension $\tilde{d}$ of process $P'$ satisfying the following bounds:
\begin{align}
R_F^{\mathrm{(S)}}(P,P')&\leq O(\epsilon_{\tilde{d}})\nonumber\\
R_F^{\mathrm{(S)}}(P,P')&\leq O\left(\frac{C_q}{\tilde{D}_q}\right),
\end{align}
where $\epsilon_{\tilde{d}}:=\sum_{k=\tilde{d}+1}^d\lambda_k$ and $\tilde{D}_q:=\log_2(\tilde{d})$.
\end{col}

From the bound $R_F^{(S)}(P,P')\leq R_F(\ket{P},\ketx{\tilde{P}})$ we can readily read off upper bounds on the distortion of our truncated quantum models using the results in Sec.~\ref{sec:examples}. We compare these with \emph{lower} bounds on the distortion of any truncated classical models of the same memory dimension $\tilde{d}_c$. The procedure by which these lower bounds can be determined is outlined in App.~\ref{app:classicalreduction}, and displayed alongside the quantum bounds for the cyclic random walk ($\tilde{d}_c\in\{3,5,7\}$) and Dyson-Ising chain processes ($\tilde{d}_c=2$, $L\in\{3,5,7\}$ in Figs.~\ref{fig:tcrwalksu} and \ref{fig:tdising}(a) respectively, showing significantly --by several orders of magnitude -- better performance for the quantum models. 

As a further point of comparison, we also consider the KL divergence rate (see App.~\ref{app:kl}), an analogous quantity to the statistical FDR that is commonly used in machine learning. We calculate the KL divergence rate of our truncated quantum models of the Dyson-Ising chain process, and lower bounds on the same for the truncated classical models; this also demonstrates much superior performance for the quantum models [\figref{fig:tdising}(b)]. Note that the same calculation could not be meaningfully made for the cyclic random walks due to the extreme sensitivity of the KL divergence to vanishing probabilities.


\section{Discussion}

Quantum sample states are a valuable resource in studying stochastic systems, as they are the requisite input for many quantum algorithms for sampling and analysing their properties. For stochastic processes such q-samples can be constructed piecemeal, one step at a time. Yet, this requires the use of an ancillary memory system that carries the information about the correlations in the process, and for complex processes this may require a high-dimensional memory. Here, we have introduced an approach to quantum dimension reduction, wherein this memory size may be reduced. We have provided theoretical heuristics demonstrating that when the amount of information carried by the memory is much smaller than the capacity offered by the size of its memory, a significant reduction in memory size is possible without significant distortion to the resulting q-sample. Further, we prescribed a systematic approach for performing this quantum dimension reduction, and demonstrated its efficacy with two illustrative examples: one a large-alphabet Markovian process, and the other a highly non-Markovian process, showing in both cases high fidelity reconstruction of the respective q-samples is possible with a significantly reduced memory size.

Moreover, we have linked this approach to (quantum) stochastic modelling, where we are tasked with reproducing the output statistics of a stochastic process. Prior works have shown that quantum models of complex stochastic processes often have much less stringent demands on the amount of information that must be stored in memory compared to their classical counterparts~\cite{aghamohammadi2016extreme, Garner2017Unbounded, elliott2018superior, elliott2019memory, elliott2022quantum}. Nevertheless, outside of a small number of examples with explicit constructions~\cite{elliott2020extreme, thompson2018causal, wu2023implementing, elliott2024embedding}, there are heretofore no known systematic methods for converting this into a reduced memory dimension -- or indeed, any guarantees that such a method exists in general for exact modelling. Our approach overcomes this issue by trading off a small distortion in statistics to achieve a potentially large reduction in quantum memory dimension, reducing it commensurately with the amount of information in must store. This complements a similar work that developed a `quantum coarse-graining' approach to dimension reduction in quantum modelling of continuous-time stochastic processes~\cite{elliott2021quantum}, providing also theoretical guarantees and a stronger performance than an initial foray into such for discrete-time processes~\cite{banchi2023accuracy}.

Being based on MPS truncation techniques, our results sit amongst a growing body of work revealing the broad power of tensor network-based approaches in studying stochastic dynamics and sampling problems. These include improved computational techniques for classical simulation of classical stochastic dynamics (including efficient estimation of high-variance observables)~\cite{hieida1998application, carlon1999density, carlon2001critical, johnson2010dynamical, johnson2015capturing, merbis2023efficient}, a means of efficiently representing classical and quantum hidden Markov models and general observer operator models~\cite{monras2011hidden, Kliesch2014, yang2018matrix, glasser2019expressive, adhikary2021quantum}, efficient calculation of distances between stochastic processes~\cite{yang2020measures}, in machine learning  and generative modelling (such as `Born machines')~\cite{Stoudenmire2016, han2018unsupervised, stoudenmire2018learning, Guo2018, Chen2018, Clark_2018, glasser2019expressive, liu2018differentiable, coyle2020born, kiss2022conditional, gong2022born}, and for classical simulation of memoryful open quantum system dynamics~\cite{strathearn2018efficient, cygorek2022simulation, gribben2022exact}. To our knowledge, ours is the first work to explore the use of tensor network techniques to simplify quantum simulation of classical processes.

More broadly, our approach to quantum dimension reduction for quantum sampling and quantum stochastic modelling may provide a route to leveraging quantum approaches for more efficient feature extraction~\cite{guyon2006introduction, khalid2014survey}, a task of vital importance in a world that is becoming ever-increasingly data-rich and data-intensive. We have substantiated that quantum models can in effect do more with less, requiring smaller memories to capture expressivity not possible with larger classical memories. Moreover, by reducing the quantum resources required for such tasks, we bring them increasingly into reach of quantum processors of the nearer future.

\textbf{Data Availability} -- Data and relevant code will be made available upon reasonable request.


\acknowledgements

This work was funded by the University of Manchester Dame Kathleen Ollerenshaw Fellowship, the Imperial College Borland Fellowship in Mathematics, grant FQXi-RFP-1809 from the Foundational Questions Institute and Fetzer Franklin Fund (a donor advised fund of Silicon Valley Community Foundation), and the National Research Foundation, Singapore, and Agency for Science, Technology and Research (A*STAR) under its QEP2.0 programme (NRF2021-QEP2-02-P06).

\appendix

\section{MPS Primer}
\label{app:mps}
We here provide some further discussion on MPSs to aid the reader in understanding the formalism in the main text. For a more in-depth presentation, we refer the reader to reviews such as Refs.~\cite{orus2014practical, Bridgeman, cirac2021matrix}.

A particularly useful property of MPSs (and tensor networks in general) is that they have powerful diagrammatic representations. Each tensor is represented by a node with several legs (see \figref{fig:mps}), where each leg corresponds to an index of the tensor. For example, a rank-3 tensor, such as the site matrix of an MPS, has 3 legs. When the legs of two tensor nodes join together, this corresponds to a tensor multiplication over these two indices. Any open legs correspond to the remaining indices of the overall tensor. In the case of an MPS, these legs correspond to the physical states at each site.

\begin{figure}[htbp]
    \centering
    \includegraphics[scale=0.5]{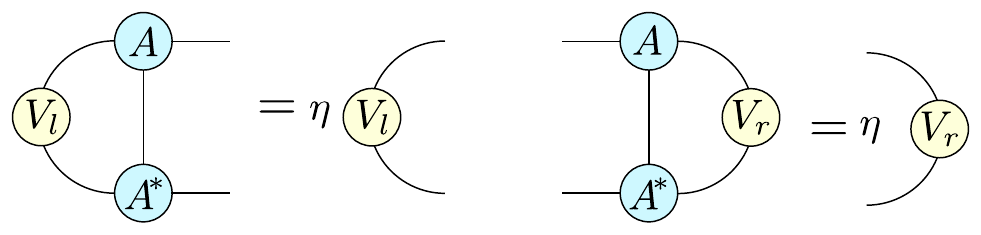}
    \caption{Diagrammatic representation of the transfer matrix of an MPS, and its left and right leading eigenvectors.}
    \label{fig:ldeig}
\end{figure}

The transfer matrix of an iMPS $\{A^x\}$ is defined as 
\begin{equation}
    \mathbb{E} = \sum_x A^x\otimes {A^x}^{*},
\end{equation}
with analogous matrices being defined for general MPS in a site-dependent manner. The left and right leading eigenvectors of $\mathbb{E}$ and denoted by $V_l$ and $V_r$ respectively, with corresponding eigenvalue $\eta$ (with $\eta=1$ for a normalised MPS), as shown in \figref{fig:ldeig}. 

A quantum state does not have a unique MPS representation; any transformation of the site matrices $A^{x,[l]}\to M^{[l]}A^{x,[l]}(M^{[l+1]})^{-1}$ for invertible matrices $\{M^{[l]}\}$ will correspond to the same state. We can use such transformations to define \emph{canonical forms} of MPSs. For an (i)MPS in left canonical form $V_l$ is the identity matrix, while for right canonical form $V_r$ is the identity matrix. We denote the corresponding site matrices as $A^x_l$ and $A^x_r$ respectively. A generic MPS can be brought into the corresponding left (right) canonical form by choosing $M$ ($M^{-1}$) to be $W_l$ ($W_r$) that satisfy
\begin{equation}
        W_l^{\dagger}W_l = V_l \quad W_rW_r^{\dagger} = V_r
\end{equation}

These can also be used to define the mixed gauge tensors of an iMPS $\{A_c, C\}$, where
\begin{equation}
\label{eq:mixedgaugeapp}
    A_c := W_l A W_r \quad C := W_l W_r.
\end{equation}
We can readily convert between these forms using $A_c = A_l C = C A_r$.

These canonical forms can be decomposed into a product of two tensors $\{\Gamma, \lambda\}$ as follows:
\begin{equation}
    A_l = \lambda\Gamma, \quad A_r = \Gamma\lambda.
\end{equation}
$\Gamma$ is a rank-3 tensor, which, like the $A$ carries a physical index corresponding to the state at the corresponding site. Meanwhile, $\lambda$ is a diagonal matrix, whose diagonal entries correspond to the (ordered) Schmidt coefficients of the (i)MPS. It can further be shown that the left and right canonical forms satisfy the completeness relations:
\begin{equation}
    \sum_x (A^x_l)^\dagger A^x_l = \mathbb{I},\quad \sum_x A^x_r(A^x_r)^\dagger=\mathbb{I}.
\end{equation}
Thus, the site matrices of the left canonical form can be interpreted as Kraus operators of a quantum channel; indeed, it is this channel that can be used to sequentially generate the MPS~\cite{Schon2005}.

\section{Proof of Theorem \ref{thm:compbound}}
\label{app:compbound}

Our proof of Theorem \ref{thm:compbound} draws heavily upon the proof of Lemma 1 in Ref.~\cite{Verstraete2006}. Specifically, in the course of said proof it is shown that given a state $\ket{\psi^L_d}$ with an MPS representation of bond dimension $d$ and $L$ sites, there exists a state $\ketx{\psi^L_{\tilde{d}}}$ with an MPS representation of bond dimension $\tilde{d}$ and $L$ sites formed from discarding the $d-\tilde{d}$ smallest Schmidt coefficients of $\ket{\psi^L_d}$ at each bond that satisfies 
\begin{equation}
|1-\brkt{\psi^L_d}{\psi^L_{\tilde{d}}}|\leq\sum_{l=1}^L\epsilon_{\tilde{d}}^{[l]},
\end{equation}
where $\epsilon_{\tilde{d}}^{[l]}$ is the sum of the discarded Schmidt coefficients at the $l$th bond. Using that $|\brkt{\psi^L_d}{\psi^L_{\tilde{d}}}|\leq 1$, this can be rearranged as
\begin{equation}
\label{eq:olboundfm}
|\brkt{\psi^L_d}{\psi^L_{\tilde{d}}}|\geq 1 - \sum_{l=1}^L\epsilon_{\tilde{d}}^{[l]}.
\end{equation}

Consider now the case where $\ket{\psi^L_d}$ and $\ketx{\psi^L_{\tilde{d}}}$ correspond to (purified) $L$-length segments of an infinite chain, such that $\epsilon_{\tilde{d}}^{[l]}$ is the same at every bond. We can then simplify \eqref{eq:olboundfm} to
\begin{equation}
\label{eq:overlapbound}
|\brkt{\psi^L_d}{\psi^L_{\tilde{d}}}|\geq 1 - L\epsilon_{\tilde{d}}.
\end{equation}

From the right canonical form of the untruncated iMPS, we can express the site matrices of the two iMPSs as $\{A_r^x\}$ and $\{\tilde{A}_r^x=A_r^x\Pi_{\tilde{d}}\}$, where $\Pi_{\tilde{d}}$ is the projection operator with support on the $\tilde{d}$ largest Schmidt coefficients. Let $\bar{\mathbb{E}}_r:=\sum_x A_r^x\otimes \tilde{A}_r^x$ be the corresponding mixed transfer matrix, which can be expressed in terms of its eigenvalues $\{\mu_j\}$ (labelled in descending order) and left (right) eigenvectors $\bra{l_j}$ ($\ket{r_j}$):
\begin{equation}
\bar{\mathbb{E}}_r=\sum_j \mu_j \ket{r_j}\bra{l_j}.
\end{equation}

We can then express
\begin{align}
\label{eq:overlaptrans}
\brkt{\psi^L_d}{\psi^L_{\tilde{d}}}&=\bopk{\Lambda}{\bar{E}_r^L}{I}\nonumber\\
&=\sum_j\mu_j^L\brkt{\Lambda}{r_j}\brkt{l_j}{I},
\end{align}
where $\ket{\Lambda}:=\sum_{j=1}^{\tilde{d}}\lambda_j\ket{j}\ket{j}$ is a vector of Schmidt coefficients for the iMPSs, and $\ket{I}:=\sum_{j=1}^d\ket{j}\ket{j}$. We will now assume that the largest eigenvalue $\mu_0$ is non-degenerate for ease of notation, noting that the following can be readily adapted for the degenerate case. Combining \eqref{eq:overlapbound} and \eqref{eq:overlaptrans} we have
\begin{equation}
1-L\epsilon_{\tilde{d}}\leq |\mu_0^L \brkt{\Lambda}{r_0}\brkt{l_0}{I}|+O(|\mu_1|^L).
\end{equation}
Thus,
\begin{align}
-&\log_2(|\mu_0^L\brkt{\Lambda}{r_0}\brkt{l_0}{I}|+O(|\mu_1^L|))\nonumber\\
&=-L\log_2(|\mu_0|) -\log_2\left(|\brkt{\Lambda}{r_0}\brkt{l_0}{I}|+O\left(\left(\frac{\mu_1}{\mu_0}\right)^L\right)\right)\nonumber\\
&\leq -\log_2(1-L\epsilon_{\tilde{d}}).
\end{align}

Finally, we use that the QFDR can be expressed in terms of the leading eigenvalue of $\bar{\mathbb{E}}_r$:
\begin{align}
R_F&=-\frac{1}{2}\log_2(|\mu_0|)\nonumber\\
&\leq-\frac{1}{2L}\log_2(1-L\epsilon_{\tilde{d}})\nonumber\\
&\quad+\frac{1}{2L}\log_2\left(|\brkt{\Lambda}{r_0}\brkt{l_0}{I}|+O\left(\left(\frac{\mu_1}{\mu_0}\right)^L\right)\right)\nonumber\\
&=-\frac{1}{2L}\log_2(1-L\epsilon_{\tilde{d}})+O\left(\frac{1}{L}\right)\nonumber\\
&\leq\frac{\epsilon_{\tilde{d}}}{2(1-L\epsilon_{\tilde{d}})\ln2}+O\left(\frac{1}{L}\right),
\end{align}
where in the last line we have used the inequality \mbox{$(x-1)/x\ln2\leq \log_2 x$}~\footnote{This inequality can be readily proven by defining the function $g(x)=1-1/x-\ln x$, noting that the maximum is at $x=1$, and then using that $g(x)\leq g(1) =0$.}. This yields the statement of the Theorem.

\section{Proof of Theorem \ref{thm:truncboundmps} and Corollary \ref{col:truncbound}}
\label{app:truncboundmps}

Our proof of Theorem~\ref{thm:truncboundmps} similarly draws upon Ref.~\cite{Verstraete2006}, mirroring elements of the proof of Lemma 2 therein.

Let us define the following family of probability distributions:
\begin{equation}
\label{eq:probfamily}
\mathcal{P}(\tilde{d},\epsilon_{\tilde{d}},h)=\{p_j|\sum_{k=\tilde{d}+1}^\infty p_k=\epsilon_{\tilde{d}},p_{\tilde{d}}=h\}.
\end{equation}
All distributions in this family are majorised by the following distribution $\{\bar{p}_j\}$:
\begin{align}
\label{eq:probmajor}
\bar{p}_1 &= 1 - (\tilde{d}-1) h - \epsilon_{\tilde{d}}\nonumber\\
\bar{p}_j &= h \quad\forall 1<j\leq \tilde{d}+\left\lfloor\frac{\epsilon_{\tilde{d}}}{h}\right\rfloor\nonumber \\
\bar{p}_f &=\epsilon_{\tilde{d}}-h\left\lfloor\frac{\epsilon_{\tilde{d}}}{h}\right\rfloor,
\end{align}
where $f=\tilde{d}+\lfloor\epsilon_{\tilde{d}}/h\rfloor$, and $\lfloor x \rfloor$ denotes the floor of $x$.

Consider an iMPS with bond dimension $d\geq 3$ and Schmidt coefficients $\lambda_j$ labelled in decreasing order. Consider also the iMPS of bond dimension $\tilde{d}$ formed by truncation of the first iMPS to its $\tilde{d}\geq 3$ largest Schmidt coefficients. Define the largest kept Schmidt coefficient $h:=\lambda_{\tilde{d}}$, and $\epsilon_{\tilde{d}}:=\sum_{k=\tilde{d}+1}^d\lambda_k$ the sum of discarded Schmidt coefficients. Noting that the Schmidt coefficients are non-negative and sum to unity, they can be interpreted as a probability distribution. Specifically, they belong to the family of distributions described in \eqref{eq:probfamily}, and are thus majorised by the distribution \eqref{eq:probmajor}. We further have the following inequalities: $\tilde{d}+\epsilon_{\tilde{d}}/h \leq d$ and $h\leq \bar{p}_1$, leading to
\begin{equation}
\frac{\epsilon_{\tilde{d}}}{d-\bar{d}} \leq h \leq \frac{1-\epsilon_{\tilde{d}}}{\tilde{d}}.
\end{equation}

Using the Schur-concavity of the Shannon entropy, we thus have that
\begin{align}
H(\lambda) & \geq H(\bar{p})\nonumber\\
&=-\bar{p}_1\log_2(\bar{p}_1) - (\tilde{d}-1+\left\lfloor\frac{\epsilon_{\tilde{d}}}{h}\right\rfloor)h\log_2(h)\nonumber\\
&\quad - \bar{p}_f\log_2(\bar{p}_f)\nonumber\\
&\geq -(\tilde{d}-2)h\log_2(h)-\epsilon_{\tilde{d}}\log_2(h)\nonumber\\
&\geq -\frac{(\tilde{d}-2)\epsilon_{\tilde{d}}}{d-\tilde{d}}\log_2\left(\frac{\epsilon_{\tilde{d}}}{d-\tilde{d}}\right) - \epsilon_{\tilde{d}}\log_2(h)\nonumber\\
&\geq -\frac{(\tilde{d}-2)\epsilon_{\tilde{d}}}{d-\tilde{d}}\log_2\left(\frac{\epsilon_{\tilde{d}}}{d-\tilde{d}}\right) - \epsilon_{\tilde{d}}\log_2\left(\frac{1-\epsilon_{\tilde{d}}}{\tilde{d}}\right)\nonumber\\
&=\frac{(\tilde{d}-2)\epsilon_{\tilde{d}}}{d-\tilde{d}}[\log_2(d-\tilde{d})-\log_2(\epsilon_{\tilde{d}})]\nonumber\\
&\quad+\epsilon_{\tilde{d}}[\log_2(\tilde{d})-\log_2(1-\epsilon_{\tilde{d}})\nonumber\\
&\geq \epsilon_{\tilde{d}}\left[\frac{\tilde{d}-2}{d-\tilde{d}}\log_2(d-\tilde{d})+\log_2(\tilde{d})\right].
\end{align}

Thence,
\begin{equation}
\label{eq:entbound}
\epsilon_{\tilde{d}}\leq\frac{H(\lambda)}{\frac{\tilde{d}-2}{d-\tilde{d}}\log_2(d-\tilde{d})+\log_2(\tilde{d})},
\end{equation}
yielding the content of Theorem~\ref{thm:truncboundmps}. Corollary \ref{col:truncbound} then follows by noting that for an iMPS representation of a q-sample $C_q=H(\lambda)$, $\tilde{D}_q:=\log_2(\tilde{d})$, and loosening the bound in \eqref{eq:entbound} by noting that the first term in the denominator is non-negative.

\section{Reduced dimension classical models}
\label{app:classicalreduction}
In general it is hard to find the most accurate (i.e., lowest distortion) unifilar classical model of a process with memory size $\tilde{D}_c<D_\mu$. However, we can obtain lower bounds on their distortion by finding bounds on the distortion of a more general class of objects.

In particular, Ref.~\cite{yang2023provably} introduced \emph{pre-models}. Like unifilar models, pre-models have an encoding function that maps from pasts to memory states. However, instead of an update rule that produces the next output and updates the memory state, an $L$-length pre-model need only produce the next $L$ outputs all at once, and nothing more. It is clear that unifilar models are a subset of $L$-length pre-models. Therefore, the $L$-length pre-model with lowest distortion sets a bound on the lowest distortion achieveable by any unifilar model of the same dimension. The formal proofs of these statements are given in Ref.~\cite{yang2023provably}. 

In Ref.~\cite{yang2023provably}, it was also shown that the memory states of the best $L$-length pre-model are coarse-grainings of the causal states. Therefore, the distortion can be bounded by a brute force search over all possible coarse-grainings of the original $d$ states into $\tilde{d}$ states, calculating its minimum value for all possible assignments of transition probabilities. 

For a Markovian process, since the most recent output uniquely defines the current causal state, their $L=1$-length pre-models are define unifilar models. For cyclic random walks, the number of possibilities for coarse-graining the memory states essentially becomes intractable with increasing $N$. There is however, a very natural coarse-graining, which is essentially to mimic that of the coarse-graining done of the continuous variable in the case of a smaller $N$. That is, we consider coarse-grainings based on clustering neighbouring segments only. 

Formally, assume the circle is divided into $2^n$ segments, i.e., $\{S_0,S_1,\ldots S_{N-1}\}$ where $N=2^n$. We partition this set into $\tilde{d}$ sets of non-overlapping, equal-size contiguous segments, to define our coarse-grained states $\{\tilde{S}_j\}$, $j\in[0,\ldots,\tilde{d}-1]$. Let $K = \lfloor N/\tilde{d}\rfloor$ and $r=N - K\tilde{d}$. Then the coarse-grained state $\tilde{S}_j$ contains $K+1$ segments if $j < r$, and $K$ segments otherwise. Specifically, these states are defined as:
\begin{align}
	\tilde{S}_j &= \{S_{j(K+1)},\ldots,S_{(j+1)(K+1)-1}\}\quad\!\quad\quad j<r,\nonumber\\
	\tilde{S}_j &= \{S_{jK+r},\ldots, S_{(j+1)K+r-1}\}\quad\quad\quad\quad j\geq r.
\end{align}
We can then derive the associated transition probabilities for these coarse-grained states:
\begin{align}
P(y|\tilde{S}_j) &= \frac{P(y,\tilde{S}_j)}{P(\tilde{S}_j)}\nonumber\\
&=\frac{\sum_{S_k\in\tilde{S}_j}P(y|S_k)P(S_k)}{\sum_{S_k\in\tilde{S}_j}P(S_k)}.
\end{align}
To convert this from a pre-model to a unifilar model, the transition rule is simply that if after output $y$ the non-coarse-grained model would transition to $S_k$, then on output $y$ the model transitions to the $\tilde{S}_j$ to which $S_k$ belongs.

\begin{figure}[htbp]
    \centering
    \includegraphics[width=\linewidth]{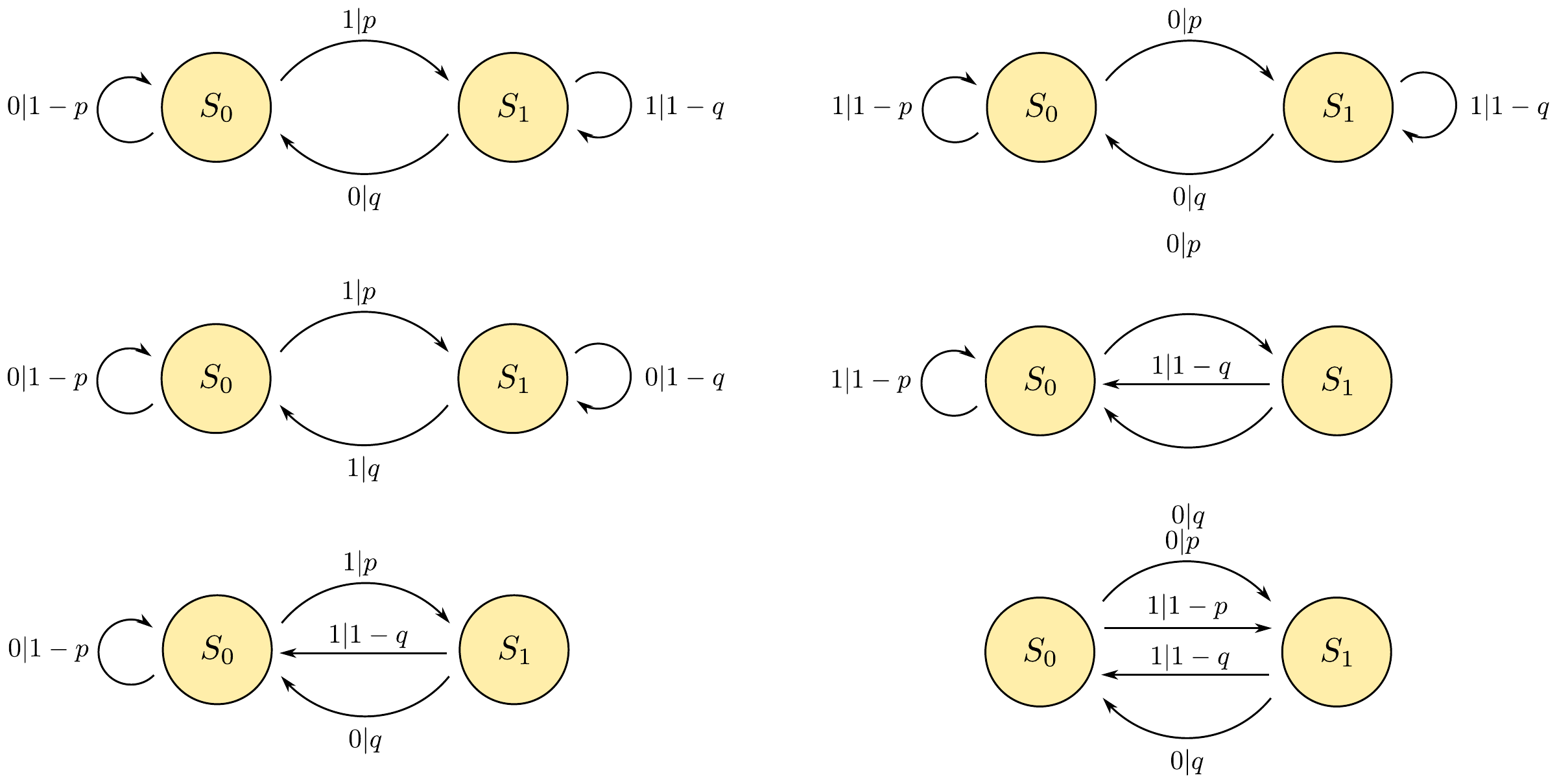}
    \caption{All topologies for two state, two symbol unifilar classical models.}
    \label{fig:2statetopo}
\end{figure}

Meanwhile, for reduced dimension classical models with $\tilde{D}_c=1$ and two outputs, the topology of possible models is sufficiently limited to allow for a brute force search directly over the set of unifilar models to obtain a tigher bound. Specifically, there are only six possible topologies, as shown in \figref{fig:2statetopo}, with each equipped with only two parameters to fully specify the transition probabilities. For the $\tilde{d}_c=2$ truncated classical models of the Dyson-Ising chain process we adopt this approach.

\section{Kullback-Leibler divergence}
\label{app:kl}

The KL divergence~\cite{cover1999elements} is a widely-used quantifier of the distance between two probability distributions $P(x)$ and $Q(x)$, defined as follows:
\begin{equation}
    \mathcal{D}_{\mathrm{KL}}(P|| Q) = \sum P(x) \log_2\left(\frac{P(x)}{Q(x)}\right).
\end{equation}
It is asymmetric in its arguments, and takes on an operational value in terms of the overhead of the cost of communicating messages drawn from $Q(x)$ using a code optimised for $P(x)$.

As with the fidelity, when applied to stochastic processes the KL divergence does not behave well, and will diverge with increasing sequence length. This is fixed by defining a KL divergence-rate that quantifies the asymptotic rate at which the KL divergence grows with increasing sequence length:
\begin{equation}
    R_{\mathrm{KL}}(P||Q) = \frac{1}{L}\mathcal{D}_{\mathrm{KL}}(P(x_{0:L})||Q(x_{0:L})).   
\end{equation}

\bibliography{References}

\end{document}